\documentclass[twocolumn]{jpsj2} 

\title{Quantum Dynamics of Spin Wave Propagation Through Domain Walls}

\author{S. Yuan$^{1}$, H. De Raedt$^{1}$ and S. Miyashita$^{2,3}$}
\inst{$^{1}$Department of Applied Physics, Materials Science Center, \\
University of Groningen, Nijenborgh 4, NL-9747 AG Groningen, The Netherlands\\
$^{2}$Department of Physics, Graduate School of Science,\\ University of
Tokyo, Bunkyo-ku,Tokyo 113-0033, Japan\\
$^{3}$ CREST, JST, 4-1-8 Honcho Kawaguchi, Saitama, Japan}

\abst{
Through numerical solution of the time-dependent Schr{\"o}dinger equation,
we demonstrate that magnetic chains with uniaxial anisotropy
support stable structures, separating ferromagnetic domains of opposite magnetization.
These structures, domain walls in a quantum system,
are shown to remain stable if they interact with a spin wave. We find that a
domain wall transmits the longitudinal component of the spin excitations only.
Our results suggests that continuous, classical spin models described by LLG
equation cannot be used to describe spin wave-domain wall
interaction in microscopic magnetic systems. }

\kword{quantum spin model, nanomagnetic, domain wall, Schr\"{o}dinger equation}

\begin{document}
\maketitle

\section{Introduction}

Wave propagation in one-dimensional magnets through a magnetic domain wall (DW) is 
an interesting topic in quantum many-body physics. A DW
separates two regions with opposite magnetization. The DW in mesoscopic
wire can be considered to be self-assembled stable nanostructures which is treated
as a kind of soliton in a continuous medium. Such structure can be created 
or annihilated by some  external action\cite{Allwood02}.
The manipulation of DW in stripes has already been proposed
as a way of storing information or even performing logic functions,
and to offer new types of electronics devices\cite{Allwood02}
in which the DW motion carries the information along a magnetic wire
of submicrometer width, with DW velocities up to thousand kilometers
per second\cite{Atkinson03}.
Recently, a direct observation of the pendulum dynamics of a DW
has been reported\cite{Saitoh04}.
The DW as a topological particle has a very small but finite mass of
$6.6\times10^{-23}$kg\cite{Saitoh04,Charppert04}.

The structure of DWs and also wave propagation in one-dimensional classical spin systems
has been studied in Refs.~\cite{NAKA74,NAKA78}.
Recently, the interaction between DWs and spin waves has attracted a lot of interest.
Hertel \textit{et al} showed
that the DW induced phase shift of spin waves in the
Landau-Lifshitz-Gilbert (LLG) model of thin, narrow strips,
is a characteristic property of such systems\cite{Hertel04}.
The value of the phase shift of spin waves passing through a DW
was found to be proportional to the angle by which the magnetization of DW
rotates in the film plane\cite{Hertel04}.
This effect might be used as a concept for a new
generation of nonvolatile memory storage and logical devices\cite{Hertel04}.

\begin{figure}[t]
\begin{center}
\includegraphics[
width=8cm
]%
{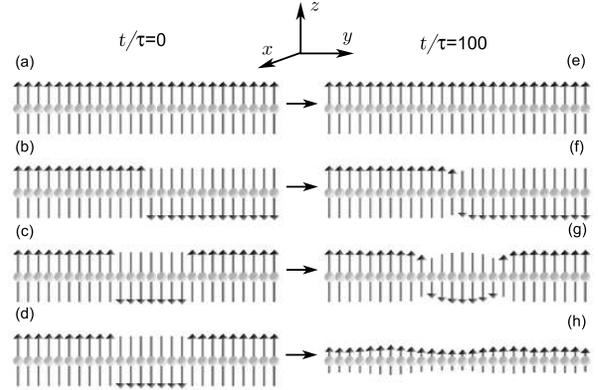}%
\caption{Left pictures (a,b,c,d): Spin configurations at time $t/\tau=0$;
Right pictures (e,f,g):
Dynamically stable spin configurations
for the Heisenberg-Ising model ($\lambda=2$),
at time $t/\tau=100$.
(e): Ferromagnetic state of the spin chain;
(f): State containing a DW at the center of the chain;
(g): State containing two DWs.
Right picture (h):
Spin configuration at time $t/\tau=100$
for the Heisenberg model ($\lambda=1$),
illustrating the instability of the initial domain wall
state (d).
}%
\label{dw_structure}%
\end{center}
\end{figure}

On the other hand, recent progress in synthesizing materials containing ferromagnetic
chains\cite{Kaji05,Mito05,Kage97,Maig00}
opens new possibilities to study the interaction between a spin wave
and a DW in a microscopic spin chain.
Furthermore, quantum spin models provide a playground to investigate
how quantum information can be transferred in quantum spin networks~\cite{BOSE03,OSBO04,CHRI04}.
But, in contrast to the nanoscale phenomena mentioned earlier,
on the atomic level, the spin dynamics is purely quantum
mechanical and in such strongly quantum fluctuating systems it must be
described by the time-dependent Schr{\"o}dinger equation (TDSE).
Then, it is of considerable interest to compare the properties
of spin wave propagation through a magnetic domain boundary in a
single spin chain with the dynamics obtained in mesoscopic system,
in which the magnetization is regarded as a classical, continuous variable.
In the nanoscale regime, the DW is defined as
the boundary of regions with opposite magnetization.
On the atomic level, a DW may be defined as a structure that
is dynamically stable under quantum mechanical motion,
the existence of which has to be confirmed.

\section{Dynamically Stable Domain Walls}

In this paper, we study the stability of DWs
and the effects of DWs on the spin wave propagation in a chain of $N$ sites
on which we place $S=1/2$ spins. We solve
the TDSE to compute the time-evolution
of the magnetization at each lattice site.
The Hamiltonian of the spin chain is given by~\cite{MATT81}
\begin{equation}
H=-J\sum_{n=1}^{N-1}(S_{n}^{x}S_{n+1}^{x}+S_{n}^{y}S_{n+1}^{y}+\lambda
S_{n}^{z}S_{n+1}^{z}), \label{H1}%
\end{equation}
where the exchange integrals $J>0$ and $\lambda J$ determine the strength of the
interaction between the $x$, $y$ and $z$ components of
spin $1/2$ operators $\mathbf{S}_n=\left(  S^{x}_n,S^{y}_n\,,S^{z}_n\right)$.
We solve the TDSE by the Chebyshev polynomial algorithm
which is known to yield extremely accurate solutions of the TDSE,
independent of the time step used~\cite{TALE84,LEFO91,IITA97,DOBR03}.
We display the results at time intervals of $\tau=\pi/5J$.
We present results for systems containing $N=26$ spins only.
We checked that simulations for $N=20$ spins (data not shown)
yield qualitatively similar results.
In our numerical work, we use units such that $\hbar=1$ and $J=1$.

First, we study the stability of DWs.
The left panel of Fig.~\ref{dw_structure} shows the spin configurations that we take
as the initial state ($t/\tau=0$) in the simulation.
All the results shown in this Letter have been obtained using open boundary conditions.
We let the system evolve in time according to the Hamiltonian
Eq.(\ref{H1}) for a long time ($t/\tau=100$)
and find that the motion generates a dynamically stable state with DW(s) (see Fig.~\ref{dw_structure}(f,g)).
The DW is defined as the boundary between regions of different magnetization
but it is not trivial that these boundaries exists in
the presence of strong quantum fluctuations.\cite{MIKE91}
Figure~\ref{dw_structure}(e,f,g) shows the dynamically stable spin configurations
obtained by starting from the corresponding configuration (a,b,c).

Whether or not quantum fluctuations destroy
the DW(s) depends on the value of the anisotropy $\lambda$.
For the model Eq.(\ref{H1}), it is well know~\cite{MATT81}
that quantum fluctuations destroy the long range order of the ground state
if $-1<\lambda<1$ (XY-like) or $\lambda=-1$ (Heisenberg antiferromagnet).
For $\lambda\ge1$ (Ising-like) the ground state exhibits long range order.
This property is reflected in the stability
of configurations that contain one or more DWs, except for $\lambda=1$.
Our numerical simulations show that
configurations with a DW are dynamically stable if $\lambda>1$.
For comparison, in Fig.~\ref{dw_structure}
we include the case $\lambda=1$, where the initial DW structure (Fig.~\ref{dw_structure}(d)) is
destroyed (Fig.~\ref{dw_structure}(h)).
Not surprisingly, the destructive effect of quantum fluctuations
can be suppressed by increasing $\lambda$.
Having studied systems with different values of $\lambda$,
we found that $\lambda=2$ is representative for the quantitative behavior of the anisotropic systems.
Therefore, in this paper, we present results for $\lambda=2$ only.
We also checked the effect of the boundary condition.
We found almost the same stable DW structures in the case of periodic boundary conditions
(results not shown).

\section{Spin Wave Propagation}

We use the configuration at $t/\tau=100$ as the initial configuration 
to study the spin wave dynamics. We generate a spin 
wave excitation by rotating the left most spin ${\bf S}_{1}$ 
in Fig.~\ref{dw_structure}(e,f,g). For reference,
we also consider the dynamics of the ferromagnet (see Fig.~\ref{dw_structure}(a,e)).
This case without DW  can be analyzed analytically, so that it also gives check 
of precision of numerical calculation. Actually, we found very small difference between
the analytical results and numerical ones.  

\begin{figure}[t]
\begin{center}
\includegraphics[
height=3.34in,
width=3.34in
]%
{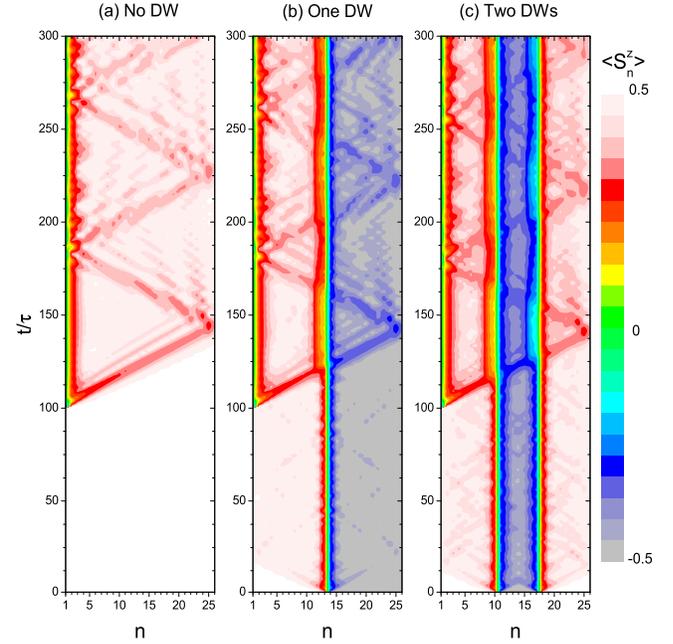}%
\caption{(color online) Time evolution of the magnetization
$\left\langle S_{n}^{z}\left(  t\right)  \right\rangle $ of the Ising-like spin chain with
$\lambda=2$. The initial configuration ($t/\tau=0$) of each panel (a,b,c) is shown
in Fig.~\ref{dw_structure}(a,b,c), respectively. At the time $t/\tau=100$, the
first spin ($n=1$) is flipped, generating a longitudinal spin wave.
}%
\label{m1_2d_20b2}%
\end{center}
\end{figure}

\begin{figure}[t]
\begin{center}
\includegraphics[
height=3.34in,
width=3.34in
]%
{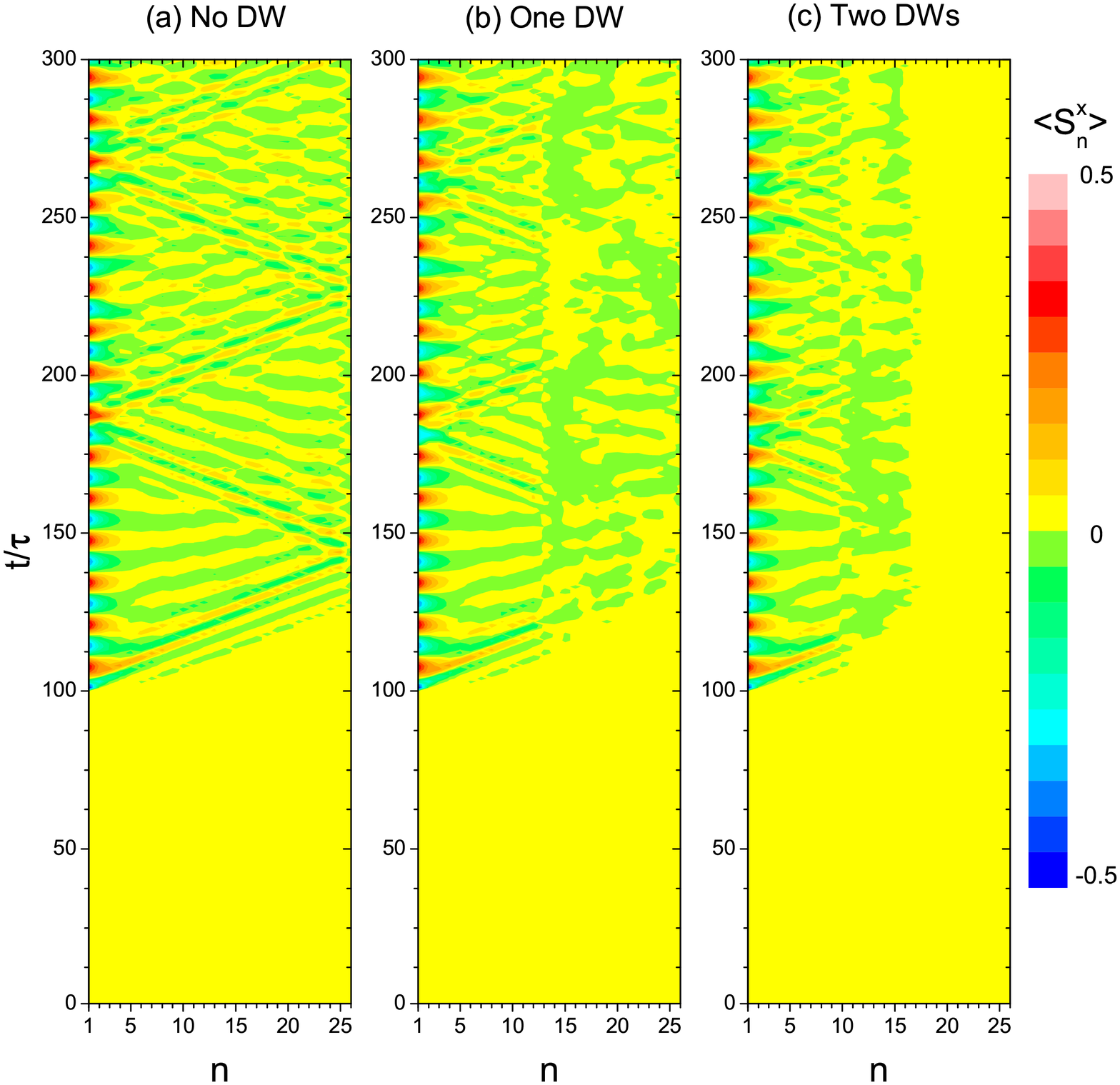}%
\caption{(color online) Time evolution of the transverse component
$\left\langle S_{n}^{x}\left(  t\right)  \right\rangle$
of the magnetization, for the same cases as those shown in
Fig.~\ref{m1_2d_20b2} except that the first spin
is rotated by $\pi/2$ about the $y$-axis (instead of flip) at the time $t/\tau=100$.}
\label{m1r_sxi2d}%
\end{center}
\end{figure}

\begin{figure}[t]
\begin{center}
\includegraphics[
height=3.34in,
width=3.43in
]%
{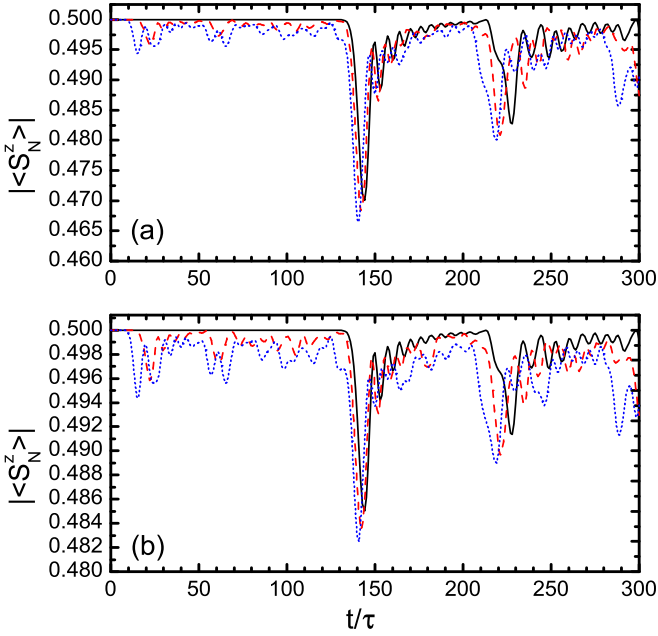}%
\caption{(color online) Time evolution of the magnetization
$\left\langle S_{N}^{z}\left(  t\right)  \right\rangle $.
(a) spin wave generated by flipping the first spin $\mathbf{S}_{1}$;
(b) spin wave generated by rotating the first spin $\mathbf{S}_{1}$
by $\pi/2$ about the $y$-axis.
Solid (black) line: No DW, corresponding to the spin configuration
Fig.~\ref{dw_structure}(e);
Dashed (red) line: one DW, corresponding to the spin configuration
Fig.~\ref{dw_structure}(f).
Because $\left\langle S_{N}^{z}\left(  t\right)  \right\rangle $ is negative
in this case, we plot the absolute value to facilitate the comparison;
Dotted (blue) line: two DWs, corresponding to the spin configuration
Fig.~\ref{dw_structure}(g).
Comparison of (a) and (b) shows that the times at which the
$\left\langle S_{N}^{z}\left(  t\right)  \right\rangle $
reaches one of the minima does not depend on method
by which the spin wave is generated.}
\label{szi20_20b2}%
\end{center}
\end{figure}

\begin{figure}[t]
\begin{center}
\includegraphics[
height=3.41in,
width=3.38in
]%
{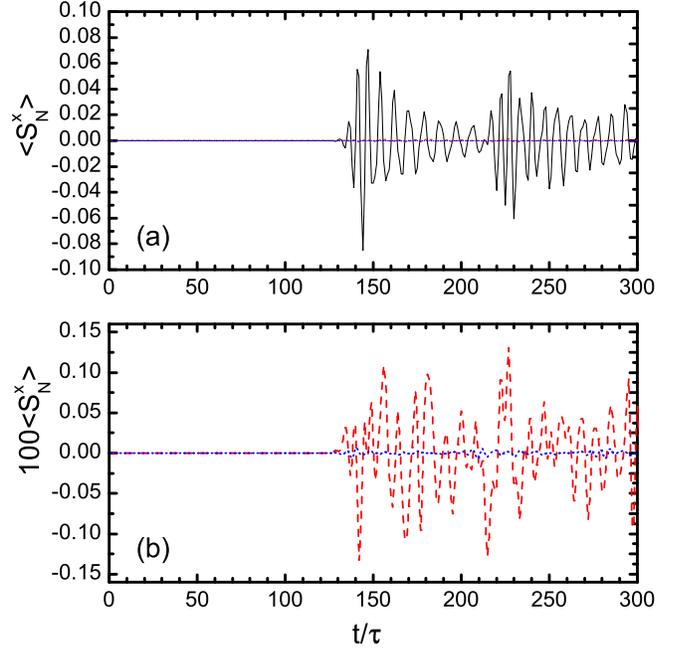}%
\caption{(color online) Time evolution of $\left\langle S_{N}^{x}\left(  t\right)
\right\rangle $ of the same system as in Fig.~\ref{m1r_sxi2d},
plotted on two different scales.
Solid (black) line: No DW, corresponding to the spin configuration
Fig.~\ref{dw_structure}(e);
Dashed (red) line: One DW, corresponding to the spin configuration
Fig.~\ref{dw_structure}(f).
Dotted (blue) line: Two DWs, corresponding to the spin configuration
Fig.~\ref{dw_structure}(g).
}%
\label{m1r_sxi20}%
\end{center}
\end{figure}

%
%

In Fig.~\ref{m1_2d_20b2}(a) we show the time evolution
of $\{\left\langle S_{n}^{z}\left(  t\right)  \right\rangle \}$ for $n=1,\ldots N$
after flipping $\mathbf{S}_{1}$, in the case of the uniform chain.
The time evolution of
$\{\left\langle S_{n}^{z}\left(  t\right)  \right\rangle \}$ for $n=1,\ldots N$
in the chain with one DW at $n=13,14$
is depicted in Fig.~\ref{m1_2d_20b2}(b),
and Fig.~\ref{m1_2d_20b2}(c) shows the results for the chain with one DW
at $n=10,11$ and another DW at $n=17,18$.
Hence, we demonstrate that even in the presence of a spin wave,
the DW structure remain stable.
In the model Eq.(\ref{H1}), the magnetization in the $z$-direction
is a conserved quantity. Hence, by flipping one or more spins
we change the total magnetization of the initial state.
The expectation value of the transverse spin components is identically zero
($\left\langle S_{n}^{x}\left(  t\right)  \right\rangle =\left\langle S_{n}%
^{y}\left(  t\right)  \right\rangle =0$ for $n=1,\ldots N$),
for all $t>0$.

From Fig.~\ref{m1_2d_20b2}, we can deduce how
the spin wave is scattered by the DW(s).
The triangular pattern in Fig.~\ref{m1_2d_20b2}(a)
merely results from the reflection of the spin flip
excitation by the other edge of the chain.
The triangular pattern is also present in Fig.~\ref{m1_2d_20b2}(b),
but the presence of the DW
causes $\left\langle S_{n}^{z}\left(  t\right)  \right\rangle $
to change sign if $n>N/2$.
Fig.~\ref{m1_2d_20b2}(b) also demonstrates that the DW itself
is extremely robust, even in systems with one spin flipped.
A similar behavior is observed for the case of two DWs
(see Fig.~\ref{m1_2d_20b2}(c)), indicating that the change
of sign at the DW is generic.

The slope of the line
in Fig.~\ref{m1_2d_20b2} from the point ($n=1,t/\tau=100$)
that connects spin 1 and spin $N$
is directly related to the velocity of the excitation.
We can estimate the time of the excitation to propagate
from site $n$ to site $m$ by analyzing the infinitely long chain.
Starting from an initial state in which we flip the spin at site $n$,
the magnetization at site $m$ is given by
\begin{eqnarray}
\langle S^z_m(t)\rangle &=&
\lim_{N\rightarrow\infty}\frac{1}{2}\left[1-|\langle n|e^{-itH}|m\rangle|^2\right],
\\
&=&\frac{1}{2}\left[1-2J^2_{m-n}(Jt)\right],
\label{Sm}%
\end{eqnarray}
where $|m\rangle$ denotes the ferromagnet state
with a flipped spin at site $m$ and
$J_{m}(x)$ is the Bessel function of the first kind of order $m$.
Although Eq.~(\ref{Sm}) is valid for the infinite chain only,
we may expect that it provides a qualitatively correct
description of the wave propagation in the finite system.
Our numerical calculations (results not shown)
demonstrate that for $N\ge16$,
the time for the excitation to travel from $n=1$ to $m=26$
agrees within $2\%$ with the first minimum of Eq.~(\ref{Sm}).

Although it is clear that the longitudinal motion of the spin that results
from the spin flip can easily propagate through the DW structures,
quantum fluctuations reduce the amplitude of the excitation
and for $t/\tau>250$ it becomes difficult to follow the excitation
in Fig.~\ref{m1_2d_20b2}(b,c).
As mentioned earlier, we could increase $\lambda$ to reduce
the quantum fluctuations but this does not change the qualitative
features that we are interested in.

Next, we study the propagation of the transverse components, that is the
$x$ or $y$ components of the expectation values of the spins.
At $t/\tau=100$, we excite the system by rotating the first spin in Fig.~\ref{dw_structure}(e,f,g) by $\pi/2$ about the $y$-axis.
After this rotation, the magnetization of spin ${\bf S}_{1}$ is parallel to the $x$-axis.
Starting from this configuration,
the time evolution will cause the first spin to rotate about the
$z$-axis (due to the presence of the neighboring spin that
is pointing in the $z$-direction).
This then generates spin waves that
contain both longitudinal
$\{ \left\langle S_{n}^{z}\left(  t\right)  \right\rangle \}$
and transverse
($\{ \left\langle S_{n}^{x}\left(  t\right)  \right\rangle \}$,
$\{ \left\langle S_{n}^{y}\left(  t\right)  \right\rangle \}$)
components.

The space-time diagram of $\left\langle S_{n}^{z}\left(  t\right)  \right\rangle $
looks very similar as Fig.~\ref{m1_2d_20b2} and therefore we do not show it.
Now, we investigate the propagation of the transverse spin waves
by considering one of the two components (the actual choice is irrelevant).
In Fig.~\ref{m1r_sxi2d}, we present results
for the time evolution of $\left\langle S_{n}^{x}\left(  t\right)
\right\rangle $ for $n=1$ to $N$.
In the Ising-like Heisenberg chain without a DW,
the transverse spin waves propagate in the same manner
as the longitudinal waves.
(compare Fig.~\ref{m1_2d_20b2}(a) and Fig.~\ref{m1r_sxi2d}(a)).
However, from Fig.~\ref{m1r_sxi2d}(b,c) it is clear that
the transverse waves do not propagate through the
DW structure but are reflected instead.

For a more quantitative study of the interaction
of DW(s) and spin waves in quantum spin chains,
we analyze in detail, the time evolution of the right-most spin.
In Fig.~\ref{szi20_20b2} we plot
$\left\langle S_{N}^{z}\left(  t\right)  \right\rangle $
as a function of time for the six cases depicted in
Figs. \ref{m1_2d_20b2} (Fig.~\ref{szi20_20b2}(a))
and \ref{m1r_sxi2d} (Fig.~\ref{szi20_20b2}(b)).

From Fig.~\ref{szi20_20b2}, we conclude that
the propagation of longitudinal spin waves in the two cases is essentially the same,
except for the amplitude.
Rotating the first spin by $\pi/2$ (instead of $\pi$ in the case
of the spin flip) about the $x$ or $y$ axis
generates waves of which the amplitude of the
longitudinal component at the site $N$ is half of that of the spin-flip case.
Using Eq.~(\ref{Sm}) and the fact that
$J_{25}(x)$ has a first maximum at $x\approx27.4$,
we find that $\langle S^z_{26}(t)\rangle$ has a minimum at $t/\tau\approx 144$.
This value is in agreement with the time
at which the numerical solution for the $N=26$ chain exhibits a first dip
(see black (solid) line in Fig.~\ref{szi20_20b2}(a)).
A first conclusion from this analysis is that
the qualitative aspects of the interaction
of the longitudinal spin wave excitation and the DW(s)
does not depend on the transverse components of the spin wave.

Fig.~\ref{szi20_20b2} also clearly shows
that the presence of a DW increases the speed
at which the excitation travels through the DW.
Comparing the curves for the system without DW,
one DW, and two DWs,
we conclude that the solid curve lags behind with respect to the dashed curve,
and the dashed curve lags behind with respect to the dotted curve.
Thus, the longitudinal component of the
spin wave excitation is shifted forward as it passes a DW.

Fig.~\ref{m1r_sxi20} shows the time evolution of
$\left\langle S_{N}^{x}\left(  t\right) \right\rangle $.
In contrast to the longitudinal component (see Fig.~\ref{szi20_20b2}),
the maximum amplitude of the transverse signals
strongly depend on the presence of DW(s)
in the system (note the difference in scale between
Fig.~\ref{m1r_sxi20}(a) and Fig.~\ref{m1r_sxi20}(b)).
Thus, in the quantum system, the reflection of the
transverse spin wave excitation is significantly larger
than the reflection of the longitudinal component.


\section{Discussion and Summary}

Finally, we point out the difference between the continuous model for mesoscale magnetic
systems and the present lattice model.
In the former, DWs exist as rotation of the spins according to a soliton structure,
while in the microscopic quantum system, there is no structure in the transverse spin component
and a DW is defined as a dynamically stable structure of the longitudinal components.
We found that such DWs exists for $\lambda>1$ whereas for
$\lambda\le1$ they are unstable.
We also studied spin wave propagation and found that the longitudinal components
of the spin wave speed up when they cross a DW. 
The transverse components of the spin wave are almost totally
reflected by the DW, but this characteristic
feature of the microscopic quantum chain 
is not found in mesoscopic magnetic system, where
the transverse components crosse a DW without reflection
and with a phase shift of $\pi/2$ \cite{Hertel04}.
It should be noted that the system described by the LLG equation
is fundamentally different from the system that we consider in this Letter.
The former treats the magnetic system in the mesoscopic regime
as a classical, continuous medium, whereas the present study
treats the magnetic system as a microscopic, quantum mechanical system.
Which of these two approaches is the most suitable description
obviously depends on the specific material.
The change of behavior from mesoscopic to microscopic
may become important as bottom-up chemical synthesis
is providing new ways for further down-sizing of the magnets.

\end{document}